%% file: Toolchain_paper.tex
\documentclass[runningheads]{llncs}

\usepackage[utf8]{inputenc}
\usepackage{ulem}
\usepackage{verbatim}

\usepackage{makeidx}
\usepackage{subcaption}
\usepackage[english]{babel}
\usepackage{graphicx}
\usepackage{amsmath,amsfonts,amssymb}
\usepackage{amstext}
\usepackage[mathscr]{eucal}
\usepackage{bm}
\usepackage{url}
\usepackage{pifont}
\usepackage{calc}
\usepackage{float}
\usepackage{tabularx} 
\usepackage{latexsym}
\usepackage{paralist}
\usepackage{xspace}
\usepackage{cancel}
\usepackage{xcolor}
\usepackage{lineno}
\usepackage{caption}
\usepackage{array}
\usepackage{multirow}
\DeclareGraphicsExtensions{.eps,.jpg,.png,.pdf}
\usepackage{amstext}
\usepackage{pifont}
\usepackage{colortbl}
\usepackage{amsmath}
\usepackage{algorithm}
\usepackage[noend]{algpseudocode}
\usepackage[pagebackref=false,bookmarks=false]{hyperref}
\usepackage{appendix}
\usepackage{float}
\usepackage{graphics}
\usepackage{subcaption}
\usepackage{xcolor}
\usepackage{xspace}
\usepackage{hyperref}
\hypersetup{
    colorlinks=false,
    linktoc=all
}
\usepackage{appendix}

\usepackage{enumitem}
\usepackage{longtable}
\usepackage{todonotes}
\usepackage{subcaption}

\usepackage{multirow,tabularx}

\usepackage{graphicx}
\newcommand\norm[1]{\left\lVert#1\right\rVert}

\begin{document}
\title{Malware Analysis with Symbolic Execution and Graph Kernel}

%
%

\author{Charles-Henry Bertrand Van Ouytsel\inst{1} \and Axel Legay\inst{1}}

%
\authorrunning{C-H. Bertrand Van Ouytsel and A. Legay}
%
\institute{INGI, ICTEAM, Université Catholique de Louvain,\\ Place Sainte Barbe 2, LG05.02,01, 1348 Louvain-La-Neuve, Belgium\\\email{\{charles-henry.bertrand,axel.legay\}@uclouvain.be}}
\maketitle 
\begin{abstract}
Malware analysis techniques are divided into static and dynamic analysis. Both techniques can be bypassed by circumvention techniques such as obfuscation. In a series of works, the authors have promoted the use of symbolic executions combined with machine learning to avoid such traps. Most of those works rely on natural graph-based representations that can then be plugged into graph-based learning algorithms such as Gspan.  There are two main problems with this approach. The first one is in the cost of computing the graph.  Indeed, working with graphs requires one to compute and representing the entire state-space of the file under analysis. As such computation is too cumbersome, the techniques often rely on developing strategies to compute a representative subgraph of the behaviors. Unfortunately, efficient graph-building strategies remain weakly explored. The second problem is in the classification itself. Graph-based  machine learning algorithms rely on comparing the biggest common structures. This sidelines small but specific parts of the malware signature. In addition, it does not allow us to work with efficient algorithms such as support vector machine.  We propose a new efficient open source toolchain for machine learning-based classification.  We also explore how graph-kernel techniques can be used in the process. We focus on the 1-dimensional Weisfeiler-Lehman kernel, which can capture local similarities between graphs. Our experimental results show that our approach outperforms existing ones by an impressive factor.
\keywords{Malware Analysis \and Symbolic Execution \and Malware Classification}
\end{abstract}

\input{Introduction}
\input{Background}

\input{Components_toolchain}

\input{Experiments}

\input{Conclusion}

\paragraph{Acknowledgments.} Charles-Henry Bertrand Van Ouytsel is an FRIA grantee of the Belgian Fund for Scientific Research (FNRS-F.R.S.). We would like to thank Cisco for their malware feed and VirusTotal for giving us access to their API.

\bibliographystyle{splncs04}
\bibliography{refs}

\end{document}

%% file: Introduction.tex
\section{Introduction}

According to the independent IT security institute AV-Test~\cite{StatSecu}, the number of malware infections has increased significantly over the last ten years, reaching a total of 1287.32 million in 2021. With approximately 450 000 new malware every day, companies spend on average $2.4$ millions dollars~\cite{budgetCyberCrime} on defenses against such malicious software. For this reason, effective and automated malware detection and classification is an important requirement to guarantee system safety and user protection.

Most malware classification approaches are based on the concept of signature and signature detection. A malware signature, which is often built manually, represents the DNA of the malware\cite{FLBGG15,KV15,biondi2018tutorial}. Consequently, deciding whether a binary file contains a specific malware boils down to checking whether the signature of such malware is present in the binary. The simplest type of signature is the syntactic signature, i.e., signatures based on syntactic properties of the malware binaries (length, entropy, number of sections, or presence of certain strings). Alternatively, behavioral signatures are based on the malware's behavioral properties (interaction with the system and its network communications).

Different types of signature give rise to different malware classification approaches. In static malware analysis approaches, the classification boils down to detecting the presence of a given static signature  directly in the binary that has been disassembled. This signature often boils down to a sequence of characters\cite{GSHC09}. The two main advantages of this approach is that it is fast and does not require executing the malware. On the other hand, static signatures are very sensitive to obfuscation techniques that modify the binary code to change its syntactic properties \cite{moser2007limits}. An illustration of those limitations is given in \cite{biondi2018tutorial,said2018detection} where the authors show the approach is not robust to variants of the MIRAI malware.

Another classification approach is that of dynamic analysis, which executes the malware and observes if its effect on the system corresponds to some behavioral signature \cite{MCNM12,ZQW20}. This approach is based on the fact that a static obfuscation does not modify the behavior of the malware and therefore has no influence on the classification of a behavioral signature. To avoid infecting the analyst’s system and to prevent the malware from spreading, the malware is commonly executed in a sandbox. Unfortunately, malware can implement sandbox detection techniques to determine whether they are being executed in a sandbox. As dynamic analysis is limited to one execution, a malware can pass detection by avoiding exhibiting malicious behavior \cite{ANSB20}. More information on static and dynamic malware analysis can be found in the following tutorial \cite{biondi2018tutorial}.

Aware of those limitations, several authors have proposed using some exploration techniques coming from the formal verification areas. This includes symbolic execution \cite{DBTMFPM16,God12,CGSS14}, a technique that explores possible execution paths of the binary without either concretizing the values of the variables or dynamically executing the code. As the code exploration progresses, constraints on symbolic variables are built and system calls tracked. A satisfiability-modulo-theory (SMT) checker is in charge of verifying the satisfiability of the collected symbolic constraints and thus the validity of an execution path. 

The advent of symbolic execution has led to the development of a new set of machine learning-based fully automatised malware classification methods. Those continue and extend the trend of applying machine learning to malware classification \cite{macedo2013mining,UAB17}. In particular, in \cite{said2018detection} the authors have proposed combining symbolic execution with Gspan \cite{yan2002gspan}, a machine learning algorithm that allows us to detect the biggest common subgraphs between two graphs. In its training phase, the algorithm collects  binary calls via symbolic analysis. Such calls are then connected in a System Call Dependency Graph (SCDG), that is a graph that abstracts the flow of information between those calls. Gspan can compute the biggest common subgraphs between malware of a given family. Those then represent the signature for the family. In its classification phase, the approach extracts the SCDG from the binary and compares it with each family's signature. In addition to being fully automatised, the approach has been shown to be more efficient than classical static and dynamic analysis approaches on a wide range of case studies. 

Unfortunately, the above-mentioned approach has several limitations. The first one is that it depends on the efficiency of the symbolic analysis engine. The second one is that SCDGs are built as an abstraction of the real behavior of the binary. In particular, the approach will connect two calls that have the same argument even though those calls may be from different function. Such a choice, which is motivated by efficiency reasons, may lead to a crude over-approximation of the file's behavior and hence to misclassification. 

Relying on the biggest common subgraphs may exclude important but isolated calls that are specific to the malware. In addition, using graphs poses a particular challenge in the application of traditional data mining and machine learning approaches that rely on vectors.  To surmount those limitations, the authors in \cite{puodzius2021accurate} proposed using a graph kernel \cite{NV21},  which can be intuitively understood as a function measuring the similarity of pairs of graphs. In their work, the authors used the approach in a non-supervised process. However, such kernel can be plugged into a kernel machine, such as a support vector machine. Results in \cite{puodzius2021accurate} show that the graph kernel outperforms Gspan in terms of accuracy. Unfortunately, the kernel used in \cite{puodzius2021accurate} still implicitly relies on detecting the biggest commonalities between graphs. Consequently, individual important calls are still out of its scope.

Our paper makes several contributions to improving symbolic analysis-based malware classification. The first contribution consists in a flexible and open source implementation of a malware analysis toolchain based on \cite{Angr} (available here~\cite{git_toolchain}). In addition to obtaining better performances, the flexibility of the new implementation allows us to plug in and compare various classification algorithms and symbolic execution strategies. In particular we develop and compare several efficient resource-based strategies that enable us to build compact but more informative SCDGs than those in \cite{said2018detection,puodzius2021accurate}. The approach is able to distinguish more SCDGs and hence obtain a finer grain in both training and classification processes. Another important contribution of this paper is the comparison of the Weisfeiler-Lehman Kernel \cite{shervashidze2011weisfeiler} with other classifiers. Such a graph kernel is capable of comparing the graph's local small structures by a ingenious relabelling of its vertices. Finally, a major contribution of the paper is a series of experimental results showing that our approach outperforms those in \cite{said2018detection} and in \cite{puodzius2021accurate} when being used in a supervised context. 

%% file: Background.tex
\section{On Graph Comparison for Malware Analysis}
\label{sec:background}

This section briefly introduces several notions related to graphs. It also outlines the limits of graph-based representation in malware analysis and advantages of graph kernels. 

A graph G is defined as a pair $(V,E)$, where $V$ is a set of Vertices and $E$ a set of edges such that $\{\{u,v\} \subseteq V | u \neq v\}$. The set of edges and vertices of $G$ are given by $E(G)$ and $V(G)$, respectively. We also consider labelled graphs where a label function $l: V(G)\rightarrow \Sigma$ assigns a label from $\Sigma$ to each vertex of $G$. We use $l(v)$ to denote the label of vertices $v$. A graph G' = (V',E') is a \textit{subgraph} of G=(V,E) if $V' \subseteq V$ and $E' \subseteq E$.

\begin{figure}[!ht]
	\centering
	\includegraphics[scale=0.30]{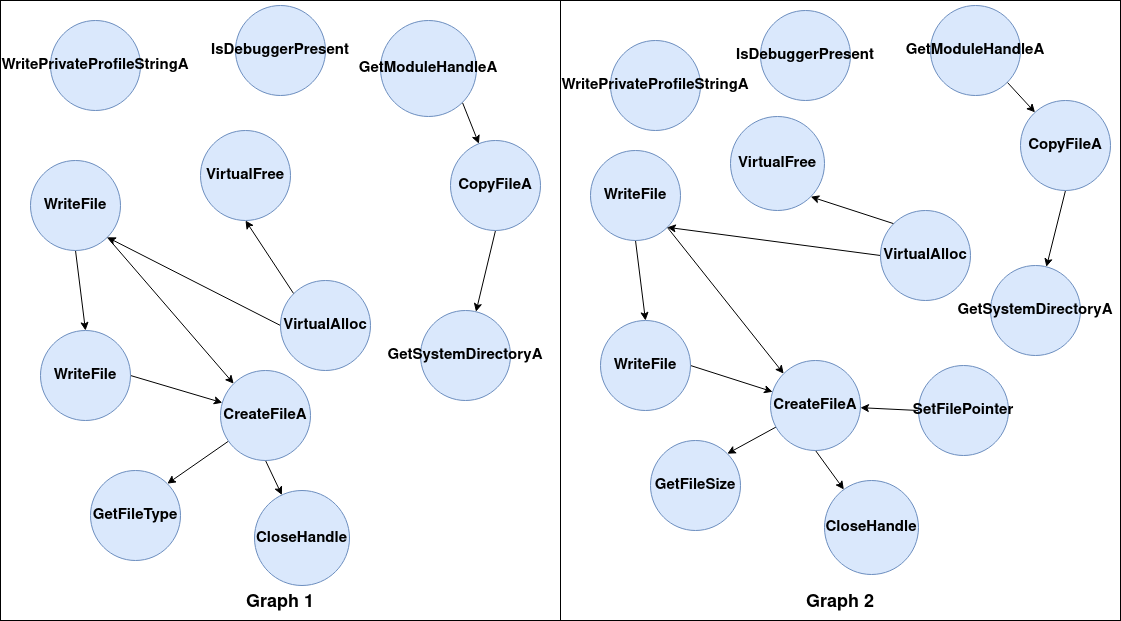}
	\caption{An example of malware behavior graphs with high similarities that are not isomorphic}\label{fig:Example_graph}
\end{figure}

In this work, we are particularly interested in applying graph comparison to extract and compare malware signatures represented by SCDGs. In particular, graph isomorphism is considered to be a powerful tool that allows us to detect structural similarities between graphs that may not be identical. Two unlabelled graphs G and H are said to be isomorphic ($G \simeq H$) if there exists a bijection $\phi: V(G) \rightarrow V(H)$ such that $(u,v) \in E(G)$ if and only if (1) $(\phi(u),\phi(v)) \in E(H)$ (for all $u,v \in V(G)$), and (2) $l(v)=l(v')$ for each $(v,v')\in \phi$. There exists a wide range of graph similarity measures. This includes, e.g., subgraph isomorphism used to compute the largest common subgraph. Checking graph isomorphism is known to be NP. Moreover, reducing the comparison of two graphs to checking their isomorphism is known to be restrictive as it requires both graphs to have same structure. This situation is rarely encountered when comparing (classes of) malware. The situation is illustrated in Figure~\ref{fig:Example_graph}, where two malware from the same family are considered to be different since Vertex $SetFilePointer$ cannot be covered by an isomorophic relationship. 

In order to leverage this problem, the authors in \cite{said2018detection} proposed an approach based on common subgraph comparison. More precisely, the authors use Gspan, which is a popular algorithm for frequent graph-based pattern mining. Given a set of graphs $\mathbb{G}$ and a desired support $min\_supp$, Gspan (whose pseudo-code is given in Figure\ref{alg:Gspan_algo}) tries to extract all subgraphs present at least in $min\_supp$ graphs of $\mathbb{G}$. If $\mathbb{G}$ represents a set of malware from the same family, the set of common subgraphs represents their signatures.  

\begin{minipage}[t]{0.45\textwidth}
\begin{algorithm}[H]
    \centering
    \caption{Gspan algorithm}\label{alg:Gspan_algo}
    \footnotesize
    \begin{algorithmic}[1]
        \State \text{In: A set of graphs:$\mathbb{G}$}
        \State \text{Out: A set of common subgraphs $\mathbb{S}$}
        \State \text{Sort labels in $\mathbb{G}$ by their frequency}
        \State \text{Remove infrequent vertices/labels}
        \State \text{Relabel remaining vertices and edges}  
        \State{$\mathbb{S}^1\gets$ all frequent 1-edge graphs in $\mathbb{G}$}
        \State{Sort $\mathbb{S}^1$ in lexicographical order}
        \State{$\mathbb{S} \gets \mathbb{S}^1$}
        \For{each edge $e \in \mathbb{S}^1$ }
        \State \text{initialize $s$ with $e$}
        \State \text{Subgraph\_mining($\mathbb{G}$ ,$\mathbb{S}$, $s$ )}
        \State \text{$\mathbb{D} \gets \mathbb{D} - e$}
        \If {$|\mathbb{G}| < min\_supp$} 
        \State \text{break}
        \EndIf
        \EndFor
        \State \textbf{Return}  \text{$\mathbb{S}$}
    \end{algorithmic}
\end{algorithm}
\end{minipage}
\hfill
\begin{minipage}[t]{0.45\textwidth}
\begin{algorithm}[H]

    \caption{Main procedure $Subgrap\_mining$ of Gspan algorithm}
    \footnotesize
    \begin{algorithmic}[1]
        \If {$s \neq min(s) $} 
        \State \text{return;}
        \EndIf
        \State \text{$\mathbb{S} \gets \mathbb{S} \cup \{s\}$}
        \State \text{enumerates s in each graph in $\mathbb{G}$}
        \State \text{and count its children;}
        \For{each c; c is a child of s}
        \If {$support(c) \geq min\_supp$} 
        \State \text{s $\gets c$}
        \State \text{Subgraph\_mining($\mathbb{G}$ ,$\mathbb{S}$, $s$ )}
        \EndIf
        \EndFor
    \end{algorithmic}
\end{algorithm}
\end{minipage}

Unfortunately, relying on computing the biggest subgraphs may dismiss small but important connected components that do not belong to the biggest subgraphs. The situation is illustrated in Graph 3 of Figure~\ref{fig:methods_limits}, where important calls such as $IsDebuggerPresent$ may be ignored. An inefficient solution could be to extend the number of subgraphs. Unfortunately, when bigger graphs than in our example are involved, this approach will mostly favor a variant of the biggest connected component, as we will see in Section~\ref{sec:experiments}. In order to leverage this problem, we resort to the concept of Graph Kernels. 


\subsection{Graph kernels}
In machine learning, \textit{kernel methods} are algorithms that allow us to compare different data points with a particular similarity measure. Consider a set of data points $X$ such as $\mathbb{R}^m$ and let $k:X \times X \rightarrow \mathbb{R}$ be a function. Function $k$ is a valid kernel on $X$ if there exists a Hilbert space $\mathcal{H}_k$ and a feature map $\phi: X \rightarrow \mathcal{H}_k$ such that $k(x,y) = \langle \phi (x),\phi (y) \rangle$ for $x,y \in \mathcal{X}$, where $\langle\cdot,\cdot\rangle$ denotes the inner product of $\mathcal{H}_k$. It is known that $\phi$ exists only if $k$ is a positive semidefinite function. A well-known kernel is the Gaussian radial basis function (RBF) kernel on $\mathbb{R}^m$, $m\in \mathbb{N}$, defined as:
\begin{equation}
  k_{RBF}(x,y) = exp(-\frac{\norm{x-y}^2}{2\sigma^2})  
\end{equation}
with $\sigma$, the bandwidth parameter. Observe that RBF kernel gives an explicit definition of $\phi$. In practice, this is not always required. Indeed, algorithms such as Support Vectors Machine (SVM) use the data $X$ only through inner products between data points. Having the kernel value $k(x,y)$ between each data point is thus sufficient to build an SVM-based classifier. This approach is known as the \textit{kernel trick}~\cite{hofmann2006support}. A Gram matrix $K$, is defined with respect to a finite set of point $x_1,..,x_n \in X$. Each element $K_{i,j}$ with $i,j\in \{0,..,n\}$ represents the kernel value between pairs of points $k(x_i,x_j)$. If the Gram Matrix $K$ of Kernel $k$ is positive semi-definite for every possible set of data points, then k is a valid kernel. 

It is common for kernels to compare data points using differences between data vectors. However, the structures of graphs are invariant to permutations of their representations (i.e., ordering of edges/vertices does not influence structure and distance between graphs). This motivates the need to compare graphs in ways that are permutation invariant. Moreover, to avoid strict comparison (which would be equivalent to isomorphism), it is common to use smoother metrics of comparison, such as \textit{convolutionnal kernels}, for better generalization capabilities. Convolutionnal kernels divide structures (i.e., graphs in our case) into substructures (e.g., edges, subgraphs, paths, etc)  and then evaluate a kernel between each pair of such substructures. 
\\

\begin{figure}[ht]
	\centering
	\includegraphics[scale=0.30]{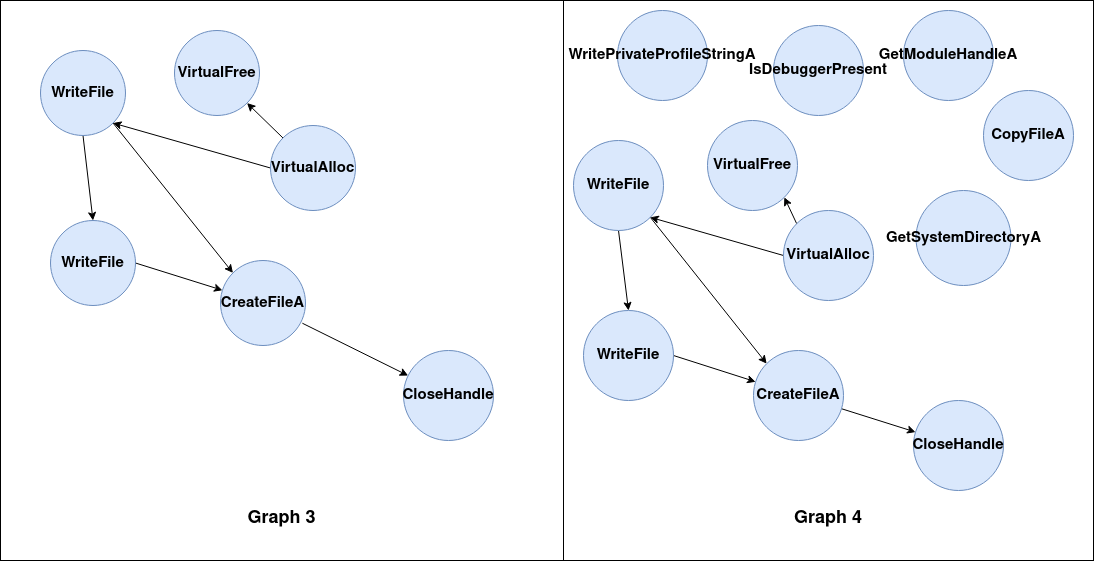}
	\caption{Graph $3$ represents the subgraph extracted with Gspan from graphs of Figure~\ref{fig:Example_graph}. Graph $4$ corresponds to the extraction with the kernel from~\cite{puodzius2021accurate}}\label{fig:methods_limits}
\end{figure}

In~\cite{puodzius2021accurate}, the authors propose a similarity metric for malware behavior graphs based on common vertices and edges.  Concretely, they define a similarity $\sigma$ between two graphs G and H as:

\begin{equation}
\sigma(G,H) = \alpha \sigma_{vertices}(G,H) + (1- \alpha) \sigma_{edges}(G,H)
\end{equation}
where $\alpha$ is the vertice-edge factor allowing to adjust weights of vertices and edges in the similarity function (set to 0.25 in the conclusion of their work). The vertice similarity is defined as:
\begin{equation}
\sigma_{vertices}(G,H) = \frac{|\mathcal{V}(G) \cap \mathcal{V}(H)|}{min(\mathcal{V}(G), \mathcal{V}(H))}
\end{equation}
 and the edge similarity as:

\begin{equation}
\sigma_{edges}(G,H) = \frac{|\mathcal{CC}_{max}(G \cap H)|}{min(|\mathcal{CC}_{max}(G)|,|\mathcal{CC}_{max}(H)|)}
\end{equation}
where $\mathcal{V}(G)$ are the set of vertices of G and $\mathcal{CC}_{max}(G)$ is the biggest connected component of G.
\newline
While this approach adds information related to all nodes labels compared to Gspan, it suffers from similar drawbacks than Gspan. Indeed, it focus on the biggest connected component, neglecting edges in other connected components. This problem is illustrated on Graph 4 of Figure~\ref{fig:methods_limits}. One can see that the kernel identifies similarities between nodes of Graph $1$ and Graph $2$. However, it ignores important edge dependencies such as \textit{GetModuleHandle}, \textit{CopyFileA}, and \textit{GetSystemDirectoryA}.\\ 

To tackle this problem, a popular approach in graph kernels is the comparison of local structure. In this approach, two vertices of different graphs are considered to be similar if they share the same labels. The two vertices are considered to be more similar if, in addition, they share similar neighborhoods (i.e., vertices with the same labels). Using this approach, Shervashidze et al.~\cite{shervashidze2011weisfeiler} introduced graph kernels based on the 1-dimensional Weisfeiler-Lehman (WL). Let $G$ and $H$ be graphs, and  $l: V(G) \cup V(H) \rightarrow \Sigma$ be a function giving their vertices labels. By several iterations $i=0,1,...$, the 1-WL algorithm computes a new label function $l_i: V(G) \cup V(H) \rightarrow \Sigma $, with each iteration allowing comparison of G and H. In the first iteration, $l_0 = l$, and in subsequent iterations,

\begin{equation}
  l_i(v) = \textit{relabel}( l_{i-1}(v), \textit{sort}(l_{i-1}(u) | u \in N(v)) )
\end{equation}
with $v \in V(G) \cup V(H)$, sort(S) returning a sorted tuple of S and function relabel(p) maps the pair p to a unique value in $\Sigma$ which is not already used in previous iterations. When the cardinality of $l_i$ equals the cardinality of $l_{i-1}$, the algorithm stops. The idea of the WL sub-tree kernel is to compute the previous algorithm for $h\ge0$
and after each iteration $i$ to compute a feature vector $\phi_i(G) \in \mathcal{R}^{|\Sigma_i|}$ for each graph G, where $\Sigma_i \subseteq \Sigma$ denotes the image of $l_i$. Each component $\phi_i(G)_{\sigma^i_j}$ counts the number of appearances of vertices labelled with $\sigma^i_j \in \Sigma_i$. The overall feature vector $\phi^{WL}(G)$ is defined as the concatenation of the feature vectors of all h iterations, i.e.,
\begin{equation}
\phi^{WL}(G) = (\phi^{0}(G)_{\sigma^0_1}, ... ,\phi^{0}(G)_{\sigma^0_{|\Sigma_0|}}, \phi^{h}(G)_{\sigma^h_1}, ..., \phi^{h}(G)_{\sigma^h_{\Sigma_h}})
\end{equation}

\begin{figure}[ht]
	\centering
	\includegraphics[scale=0.3]{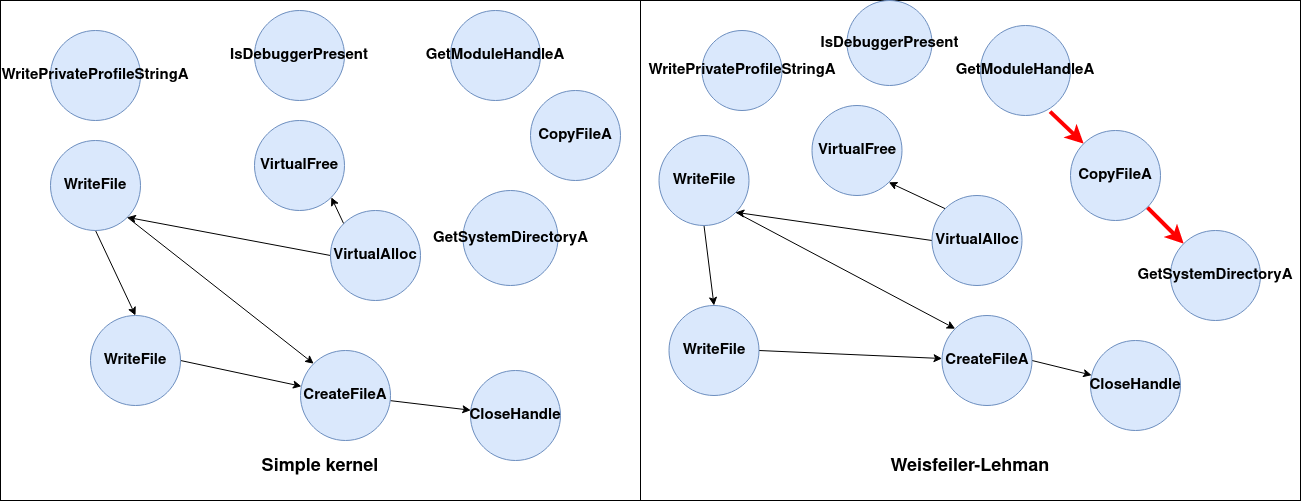}
	\caption{The figure shows that, contrary to Simple Kernel, the Weisfeiler-Lehman kernel captures all common edges between the graphs of Figure~\ref{fig:Example_graph}}\label{fig:graph_kernel_illus}
\end{figure}

Finally, to compute similarity between two different feature vectors, we apply the following formula:

\begin{equation}
    k_{WL}(G,G')=\sum_{\phi \in \phi^{WL}(G)}\sum_{\phi' \in \phi^{WL}(G')} \delta(\phi,\phi')
\end{equation}
where $\delta$ is the Dirac kernel, that is, it is $1$ when its arguments are equals and $0$ otherwise. The more labels the two graphs have in common, the higher this kernel value will be. Compared with Gspan and the kernel from~\cite{puodzius2021accurate}, this kernel also targets similarities related to all nodes and edges of the biggest subgraph but also local similarities. This is illustrated in Figure~\ref{fig:graph_kernel_illus}, where dependencies between \textit{GetModuleHandle}, \textit{CopyFileA}, and \textit{GetSystemDirectoryA} are kept in the learning process.

%% file: Components_toolchain.tex
\section{An Open Source Toolchain for Malware analysis}
\label{sec:component_toolchain}
\begin{figure}[h]
	\vspace*{-0.0cm}
	\centering
	\includegraphics[width=14.0cm]{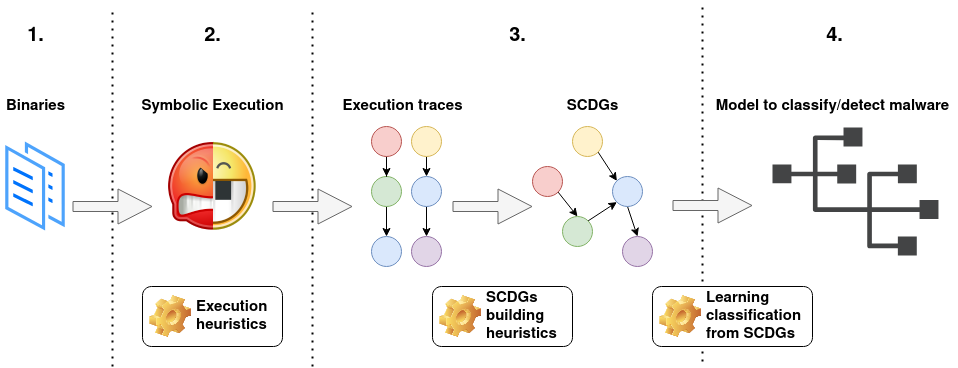}
	\caption{General view of the toolchain}\label{fig:view_toolchain}
\end{figure}

We propose an open source toolchain for malware analysis that is based on machine learning and SCDGs (available here~\cite{git_toolchain}). The toolchain, which is represented in Figure~\ref{fig:view_toolchain}, relies on the following important components: the first component consists in collecting and labelling a series of binaries from different malware families. Then, \textbf{Angr}~\cite{Angr}, a python framework for symbolic execution, is used to execute those files. The result is used to extract a SCDG for each such binary. One of the contributions of this paper will be to improve and adapt the symbolic engine to malware analysis as well as the construction of SCDGs. Those SCDGs are then used to train machine learning algorithms. If Gspan is used, the training will result in common subgraphs to represent signatures for each family. If SVM is used, a Gram matrix between all the malware programs is created. Finally, the toolchain also contains supervised classifiers. If Gspan is used, the SCDG of the new malware is compared with those of the signature of each family and the classifier retains the one with the closest distance. If SVM is used, a Gram matrix is created between all trained malware and the new malware. This matrix is then used in the SVM classification process. A main contribution of this paper is to compare those two types of classification.

\subsection{Extraction of calls}

The construction of the SCDG is based on Symbolic Execution. This approach envisages the exploration of all the possible execution paths of the binary without either concretizing the values of the variables or dynamically executing the code (i.e., the binary is analyzed statically). Instead, all the values are represented symbolically. As the code exploration progresses, constraints on symbolic variables are built and system calls tracked. A satisfiability-modulo-theory (SMT) checker is in charge of verifying the satisfiability of the collected symbolic constraints and thus the validity of an execution path. 
A wide range of tools and techniques have been developed for efficient symbolic execution analysis. Most of those techniques agree on the fact that symbolic execution still suffers from state-space-explosion and, consequently, only a finite set of symbolic paths can be explored in a reasonable amount of time. This is particularly the case with malware analysis where the classification process must be done with very limited resources. As the calls that form the SCDG are collected directly from those symbolic paths, the choice of which paths to follow will have an impact on the machine learning process.In a recent work, authors showed how SMT solving could impact performances \cite{sebastio2020optimizing,BJLS17,CCSZP0021}. In this paper, we focus on path selection strategies. The work in \cite{said2018detection} implements a Breadth-First Search (BFS) approach, that is, at each execution step all ongoing paths are explored simultaneously. This approach leads to an important growth of states and memory usage. As we have limited resources, we propose to explore one subset of paths at a time. We prioritize states from which one can explore new assembly instruction addresses of the program.  Our Custom Breadth-First Search Strategy ({\bf CBFS-Strategy}) is presented in~\ref{alg:CBFS}. The algorithm begins by taking $L$ states for exploration from the set of available states and putting them in the list $R$ of states to explore next (line $4$). It then iterates among all other available states. If it finds a state leading to an unexplored part of the code or with a shorter path of execution (line $6$), it puts it in $R$ and takes out a state with a lower priority. After going through each state, it returns $R$ to allow ANGR to perform a new execution step on $R$' states. In addition to {\bf BFS-Strategy}, we also implemented a  Custom Depth-First Search Strategy ({\bf CDFS-Stategy}), which is presented in Algorithm \ref{alg:CDFS} (the main difference with CBFS-Strategy being the condition to select successor state at Line $6$). Observe that symbolic execution with depth and breadth first search is not new. However, the implementation and evaluation of restricted versions within a tool for malware classification are. 

\begin{minipage}{0.45\textwidth}
\begin{algorithm}[H]
    \centering
    \caption{\textbf{CBFS} exploration}\label{alg:CBFS}
    \footnotesize
    \begin{algorithmic}[1]
        \State \text{Inputs: A set of states: S}
        \State \text{Limit of states: L}
        \State \text{Outputs: A set of L states: R}
        \State{$R\gets S[:L]$}
        \For{$state \in S$}
        \If {$\text{new(state.next\_ip)} |\text{state.depth} < \{s.depth | s \in R \}$} 
        \State \text{Remove s from R}
        \State \text{Add state to R}
        \EndIf
        \EndFor
        \State \textbf{Return}  \text{R}
    \end{algorithmic}
\end{algorithm}
\end{minipage}
\hfill
\begin{minipage}{0.45\textwidth}
\begin{algorithm}[H]
    \centering
    \caption{\textbf{CDFS} exploration}\label{alg:CDFS}
    \footnotesize
    \begin{algorithmic}[1]
        \State \text{Inputs: A set of states: S}
        \State \text{Limit of states: L}
        \State \text{Outputs: A set of L states: R}
        \State{$R\gets S[:L]$}
        \For{$state \in S$}
        \If{$\text{new(state.next\_ip)} | \text{state.depth} > \{s.depth | s \in R \}}$
        \State \text{Remove s from R}
        \State \text{Add state to R}
        \EndIf
        \EndFor
        \State \textbf{Return}  \text{R}
    \end{algorithmic}
\end{algorithm}
\end{minipage}

Another important challenge in symbolic execution is that of handling loops. Indeed, the condition of such loops may be symbolic. In addition, the loop may create an infinite repetitive behavior. In those situations, deciding between staying in the loop or exiting the loop remains a tricky choice that has been the subject of several works focusing on the possibilities, which include Loop-extended Symbolic Execution~\cite{saxena2009loop}, Read-Write set~\cite{rwset}, and bit-precise symbolic mapping~\cite{xu2017cryptographic}. As those approaches may be too time-consuming, we propose to reuse two intermediary heuristics from \cite{said2018detection}. The first one applies to loops whose condition contains a symbolic value. Such loops may give rise to two states at each iteration: one that exits the loop for those symbolic values that exceed the condition and one that remains within the loop for other values, with this last state being used again to iterate on the loop. We chose to stop such iteration after four steps for loops that do not contain symbolic values, since such loop may still lead to an unbounded number of behaviors, our approach consists in limiting the execution to a finite precomputed number of steps and then forcing the execution to exit the loop. 

\subsection{Creating SCDGs}

Symbolic execution allows us to obtain several paths representing executions of a given binary. Our next step is to collect the sets of calls present on each such path as well as their addresses and arguments. Those are used to build the SCDG corresponding to this binary. ).  Following \cite{said2018detection}, SCDGs are graphs where each vertex is labelled with the name of a system call; and the edges correspond to (an abstraction of) information flow between these calls. Concretely, each SCDG is built from the symbolic representation by merging and linking calls from one or more symbolic paths. 

Consider first the creation of a graph execution from one symbolic path. We consider three types of edge. In the first one, two calls are linked if they both share an argument with identical value. This is, for example, the case of two calls with the same file handler. The second link is established between two calls that both have an argument with the same symbolic value. An example is a symbolic file size returned by a call and passed to a second call added to another value. In addition, we consider that two calls can be linked if an argument of the first call is the calling address of the second one. This situation typically arises in dynamic loading of a library. We also label each edge with the index of the argument in both calls (return value of a call is given index $0$). The three-edges strategy is called {\bf SCDG-Strategy 1} and the one-edge strategy is called {\bf SCDG-Strategy 2}. Our experiments shows that {\bf SCDG-Strategy 2} loses important dependency between calls and leads to more isolated nodes in SCDGs. Indeed, this strategy suffers from two types of problem.  First, symbolic values may be modified before being passed to another call. Second, some calls used by obfuscation techniques exhibit address-arguments links. A typical example is given by \textit{GetProcAddress}, used to hide real content of the import table of PE files.

\begin{figure}[h]
	\vspace*{-0.0cm}
	\centering
	\includegraphics[width=14.0cm]{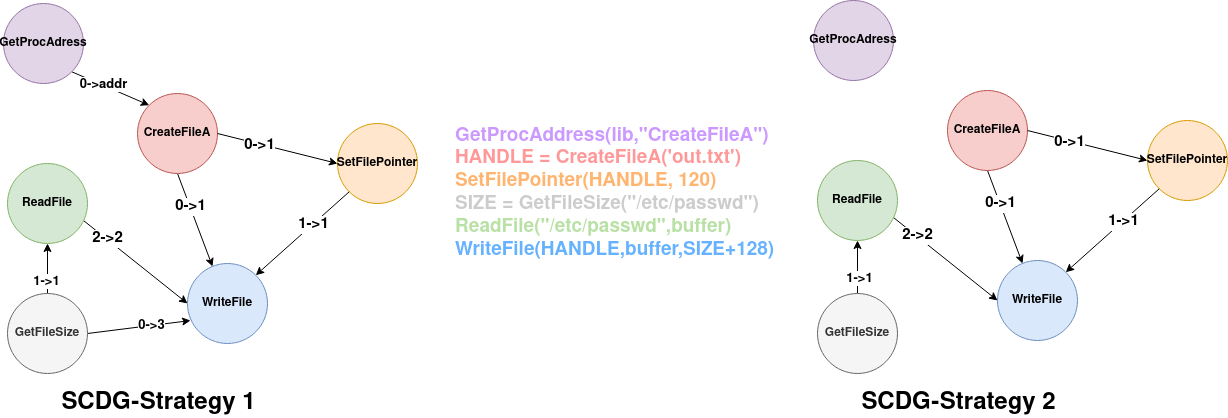}
	\caption{Illustration of a SCDG built with {\bf SCDG-Strategy 1} and {\bf SCDG-Strategy 2}}\label{fig:SCDG_example}
\end{figure}

 An example of an SCDG is given in Figure~\ref{fig:SCDG_example} with {\bf SCDG-Strategy 1} and {\bf SCDG-Strategy 2}. The program first calls \textit{CreateFile}, which returns a handle to the file with the specified filename. A vertex is thus constructed for \textit{CreateFile}. Then, a call to \textit{SetFilePointer} on the preceding file handle occurs. This leads to the creation of a new vertex (\textit{SetFilePointer}). Since  the returned argument of \textit{CreateFile} (index 0) is the same as the first argument of \textit{SetFilePointer} (index 1), an edge is added between them. Vertices \textit{ReadFile} and \textit{WriteFile} are created and linked following similar principles.

There are situations where different calls in the same execution share the same API name and occur at the same instruction address but with distinct arguments. In such situations, one may decide to merge the two calls into one single vertex. In this case, we conserve the set of arguments of the first call observed in the execution. This merge incurs a loss of precision but leads to a more compact SCDG representation. This may be of importance when one has to train the system with a large number of different types of malware. In the rest of the paper, this merging strategy is called {\bf SCDG-Strategy 3}.  Merging calls gives different advantages. First, it decreases the size of the SCDG, which may lead to better classification/detection performances. In addition, it may reduce the impact of some calls in the learning phase. An example is given with the wabot malware, which uses a hundred of calls to \textit{WriteFile} during its execution. Those calls are not part of the main actions of the malware. If they are not merged, they will constitute an important part of the signature and may have a negative impact on the training phase. On the other hand, there are situations where {\bf SCDG-Strategy 3} will merge several calls with different goals. This situation may result in losing part of the malware behavior. 

Observe that the above strategies apply to single symbolic paths only. When several symbolic paths are considered, one can decide to produce an SCDG that is composed of the disjoint union of such executions. Such a strategy is referred to as {\bf SCDG-Strategy 5} in the rest of the paper. On the other hand, {\bf SCDG-Strategy 4} consists in merging successive executions from different symbolic paths. {\bf SCDG-Strategy 5} is simplier to compute, but {\bf SCDG-Strategy 5} gives smaller graphs. According to our experimental results, {\bf SCDG-Strategy 4} may speed up the computation time by an exponential factor for families with high symbolic execution numbers.

\subsection{Creating a classification model and evaluate new samples}

The toolchain uses SCDGs to train a classifier which is used to detect and classify malware. We have implemented two classifiers. One implementation is based on Gspan and follows the idea from \cite{said2018detection}. Another one implements the graph kernel from \cite{puodzius2021accurate} and the Weisfeiler-Lehman extension we outlined in Section \ref{sec:background}. 

The classifier that uses Gspan implementation works by extracting signatures from malware families.  We obtain the signature of each family by computing the biggest subgraphs between the SCDG of each malware. In the classification phase, we compare the SCDG of new binary with those of each signature. The file belongs to the malware family whose graph is the closest to the binary's.

For the case of graph kernel, the training phase consists in computing the feature vector that corresponds to applying the algorithm in \cite{puodzius2021accurate} or the Weisfeiler-Lehman extension to each malware of the family. As explained in the background section, the algorithm produces a Gram matrix between all those vectors. This matrix represents an implicit version of the kernel. A support vector machine can then exploit this implicit representation. In the classification phase, we compute a Gram matrix between the feature vector of the binary under classification and the vectors of all malware used in the training set. Observe that, contrary to the Gspan approach, graph kernel does not require us to produce an explicit and hence all-encompassing representation of the signature of each family.

%% file: Experiments.tex
\section{Experimental Results}
\label{sec:experiments}

This section describes the methodology used to assess our toolchain's performance in both extracting SCDGs and classifying new binaries. Our evaluation set was composed of 1874 malware divided into 15 families plus 150 cleanware samples. The data set's exact composition is given in Table~\ref{table:dataset}. In terms of origins, $64$ percent of the samples used in our data set were obtained thanks to a direct pipeline between UCLOUVAIN and Cisco between April and June 2020. The remaining $36$ percent were extracted from MalwareBazaar~\cite{MalwareBazaar} between April 2020 and December 2020. Samples were labelled using AVClass~\cite{sebastian2016avclass}, a python tool to label malware samples. This tool is fed with VirusTotal reports and outputs the most likely family of each sample. To evaluate detection performance, we used 150 open source programs found online~\cite{dataset_clean}.

\begin{table}[!htb]
\centering
\begin{tabular}{|l|l|l|l|l|}
\hline
 Family & \#samples & & Family & \# samples\\ \hline
 bancteain & 91  & & remcosRAT & 476\\ 
 delf & 78  & & sfone &   32\\ 
 fickerstealer  & 44  & & sillyp2p & 269 \\ 
 gandcrab & 92 & & simbot & 126\\ 
 ircbot & 36 & & sodinokibi & 75\\ 
 lamer & 61   & & sytro & 115\\ 
 nitol & 71 & & wabot & 134\\ 
 redlinestealer & 35 & & cleanware &  150\\  \hline
\end{tabular}
\caption{Composition of the dataset}
\label{table:dataset}
\end{table}

In the rest of the section, all experiments were performed on a desktop PC with an Intel Core i7-8665U CPU (1.90GHz x 8) and 16GB RAM running Ubuntu 18.04.5. Our experimental results relied on our ability to extract SCDGs efficiently.  In all experiments, we used a timeout of ten minutes for each SCDG. Note that 20 percent of the SCDGs were computed in time while 80 percent never computed entirely. For the case of BFS-Strategy, we used the same parameters as in \cite{sebastio2020optimizing} (loop threshold of 4, unlimited number of states to explore, z3 optimization enabled). However, for CDFS-Strategy and CBFS-Strategy we imposed a limit of 10 states that could be explored simultaneously. 


\paragraph{\textbf{Environment modeling}} Proper environment modelling is a major challenge in developing efficient symbolic execution techniques.  Indeed, when we apply symbolic execution we avoid exploring/executing API call code. Indeed, performing such an operation would drastically increase the computation time~\cite{Lin17a}.  In ANGR, when a call to an external library occurs, the call is hooked to a simulated procedure called \textit{simprocedures} that will produce the symbolic outputs for the function. A simple but crude implementation of such procedure is to assume that the external function returns a symbolic value without any constraint. In such a case, \textit{simprocedures} simply returns symbolic values covering the full range of outputs given in the specification. In practice, such a solution gives good results in 26 percent of the cases. However, this solution may generate outputs that are not defined in the specification. In addition, it ignores many potential effects of the call, which include the modifications of input parameters or the number of its arguments. This may lead to incoherent executions if those parameters impact the rest of the execution (e.g., in branch choices).  We propose several improvements to fix those issues. The first one consists in restricting the ranges of outputs to those given in the specification. As an example, if the output is an integer variable that can take only four values, \textit{simprocedures}  would generate those values instead of the full range of integers. Another one concerns the case where an execution is blocked or cheated because modifications of some arguments by the external call are not performed. This happens in situations where the external call may modify some of its inputs or even some environment variables. In such case, we emulate all potential modifications with concrete values. Finally, we also take variadic functions into account. The specification of such a function is given with a fixed number of arguments. As the number of arguments may change at execution, considering this fixed number only would have an impact on the stack. To solve this problem, we explicitly count the number of arguments passed onto the function. Finally, we also use the ANGR abstraction for file systems that allows us to reuse a file in multiple systems. Observe that this improvement work must be performed for each call that causes problems. This constitutes a tremendous amount of work. To ease the life of future developers, we have constituted a \textit{simprocedures} library that is constantly enriched with new experiments and calls.

We first apply Gspan to SCDGs obtained with combinations of different strategies. Signatures are obtained by sampling randomly 30\% of the SCDGs of each family; those SCDGs constitute the training set. Other SCDGs are then classified to assess the quality of those signatures; those SCDGs constitute the test set. This process is repeated three times and performance is averaged.

\begin{table}[!htb]
\centering
\begin{tabular}{|lllll|lll|lll|lll|}
\hline
\multicolumn{5}{|l|}{\textbf{SCDG-strategy}} &
  \multicolumn{3}{l|}{{\bf BFS-Strategy}} &
  \multicolumn{3}{l|}{{\bf CBFS-Strategy}} &
  \multicolumn{3}{l|}{{\bf CDFS-Strategy}} \\ \hline
\multicolumn{1}{|l|}{\ 1 \ } &
  \multicolumn{1}{l|}{\ 2 \ } &
  \multicolumn{1}{l|}{\ 3 \ } &
  \multicolumn{1}{l|}{\ 4 \ } &
  \ 5 &
  \multicolumn{1}{l|}{Precision} &
  \multicolumn{1}{l|}{Recall} &
  $F_1$-score &
  \multicolumn{1}{l|}{Precision} &
  \multicolumn{1}{l|}{Recall} &
  $F_1$-score &
  \multicolumn{1}{l|}{Precision} &
  \multicolumn{1}{l|}{Recall} &
  $F_1$-score\\ \hline
\multicolumn{1}{|l|}{\ x} &
  \multicolumn{1}{l|}{} &
  \multicolumn{1}{l|}{\ x} &
  \multicolumn{1}{l|}{\ x} &
   &
  \multicolumn{1}{l|}{0.685} &
  \multicolumn{1}{l|}{0.566} &
  0.619 & 
  \multicolumn{1}{l|}{0.721} &
  \multicolumn{1}{l|}{0.604} &
  0.657 & 
  \multicolumn{1}{l|}{0.652} &
  \multicolumn{1}{l|}{0.609} & 0.629
   \\ \hline
\multicolumn{1}{|l|}{\ x} &
  \multicolumn{1}{l|}{} &
  \multicolumn{1}{l|}{\ x} &
  \multicolumn{1}{l|}{} &
  \ x &
  \multicolumn{1}{l|}{0.623} &
  \multicolumn{1}{l|}{0.552} & 0.585
   &
  \multicolumn{1}{l|}{0.674} &
  \multicolumn{1}{l|}{0.568} &
   0.616&
  \multicolumn{1}{l|}{0.61} &
  \multicolumn{1}{l|}{0.587} & 0.5983
   \\ \hline
\multicolumn{1}{|l|}{\ x} &
  \multicolumn{1}{l|}{} &
  \multicolumn{1}{l|}{} &
  \multicolumn{1}{l|}{\ x} &
   &
  \multicolumn{1}{l|}{0.683} &
  \multicolumn{1}{l|}{0.589} &
   0.632 &
  \multicolumn{1}{l|}{0.698} &
  \multicolumn{1}{l|}{0.555} & 0.619
   & 
  \multicolumn{1}{l|}{0.669} &
  \multicolumn{1}{l|}{0.597} & 0.631
   \\ \hline
\multicolumn{1}{|l|}{\ x} &
  \multicolumn{1}{l|}{} &
  \multicolumn{1}{l|}{} &
  \multicolumn{1}{l|}{} &
  \ x &
  \multicolumn{1}{l|}{0.609} &
  \multicolumn{1}{l|}{0.534} & 0.569
   &
  \multicolumn{1}{l|}{0.651} &
  \multicolumn{1}{l|}{0.534} & 0.586
   &
  \multicolumn{1}{l|}{0.629} &
  \multicolumn{1}{l|}{0.54} & 0.581
   \\ \hline
\multicolumn{1}{|l|}{} &
  \multicolumn{1}{l|}{\ x} &
  \multicolumn{1}{l|}{\ x} &
  \multicolumn{1}{l|}{\ x} &
   &
  \multicolumn{1}{l|}{0.684} &
  \multicolumn{1}{l|}{0.651} & \textbf{0.667}
   &
  \multicolumn{1}{l|}{0.736} &
  \multicolumn{1}{l|}{0.655} & 0.693
   &
  \multicolumn{1}{l|}{0.693} &
  \multicolumn{1}{l|}{0.639} & 0.664
   \\ \hline
\multicolumn{1}{|l|}{} &
  \multicolumn{1}{l|}{\ x} &
  \multicolumn{1}{l|}{\ x} &
  \multicolumn{1}{l|}{} &
  \ x &
  \multicolumn{1}{l|}{0.614} &
  \multicolumn{1}{l|}{0.594} & 0.614
   &
  \multicolumn{1}{l|}{0.686} &
  \multicolumn{1}{l|}{0.615} &
   0.648 &
  \multicolumn{1}{l|}{0.645} &
  \multicolumn{1}{l|}{0.586} & 0.614
   \\ \hline
\multicolumn{1}{|l|}{} &
  \multicolumn{1}{l|}{\ x} &
  \multicolumn{1}{l|}{} &
  \multicolumn{1}{l|}{\ x} &
   &
  \multicolumn{1}{l|}{0.679} &
  \multicolumn{1}{l|}{0.587} &
    0.629 &
  \multicolumn{1}{l|}{0.623} &
  \multicolumn{1}{l|}{0.465} &
   0.532 &
  \multicolumn{1}{l|}{0.68} &
  \multicolumn{1}{l|}{0.597} & 0.636
   \\ \hline
\multicolumn{1}{|l|}{} &
  \multicolumn{1}{l|}{\ x} &
  \multicolumn{1}{l|}{} &
  \multicolumn{1}{l|}{} &
  \ x &
  \multicolumn{1}{l|}{0.602} &
  \multicolumn{1}{l|}{0.556} &
   0.578 &
  \multicolumn{1}{l|}{0.587} &
  \multicolumn{1}{l|}{0.448} &
  0.508 & 
  \multicolumn{1}{l|}{0.632} &
  \multicolumn{1}{l|}{0.578} & 0.603
   \\ \hline
\end{tabular}
\centering
\caption{Results of Gspan classifier with different exploration strategies}
\label{table:res_gspan}
\end{table}

\begin{figure}[!ht]
	\centering
\includegraphics[scale=0.30]{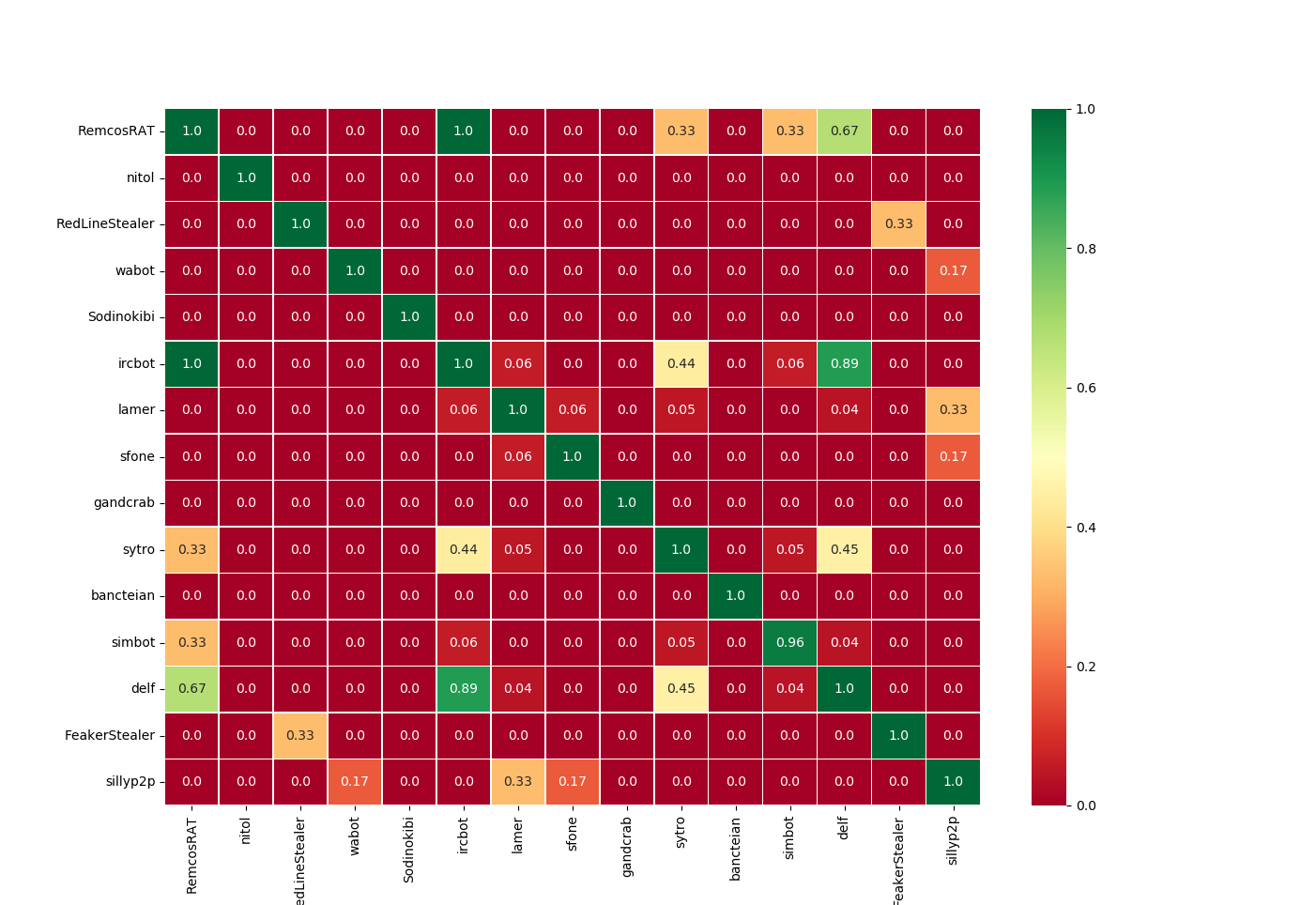}
	\caption{Similarity matrix between signatures obtained with Gspan}\label{fig:inter_sig_gspan}
\end{figure}

\begin{figure}[!h]
\begin{minipage}[h]{0.40\linewidth}
  \includegraphics[scale=0.3]{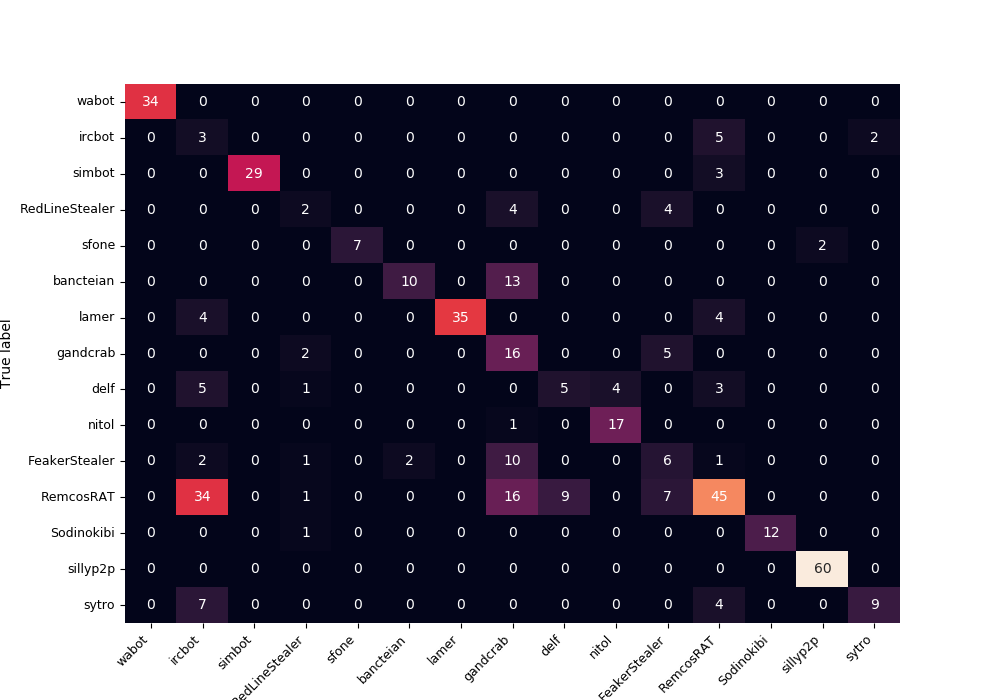}
  \caption{Confusion matrix obtained for Gspan with \textbf{CBFS-strategy} and \textbf{SCDG-strategy 2-3-4}.}
  \label{fig:conf_m_gspan}
\end{minipage}\quad \quad \quad\quad\quad \quad\quad
\begin{minipage}[h]{0.40\linewidth}
\hfill \hfill \hfill
 \includegraphics[scale=0.3]{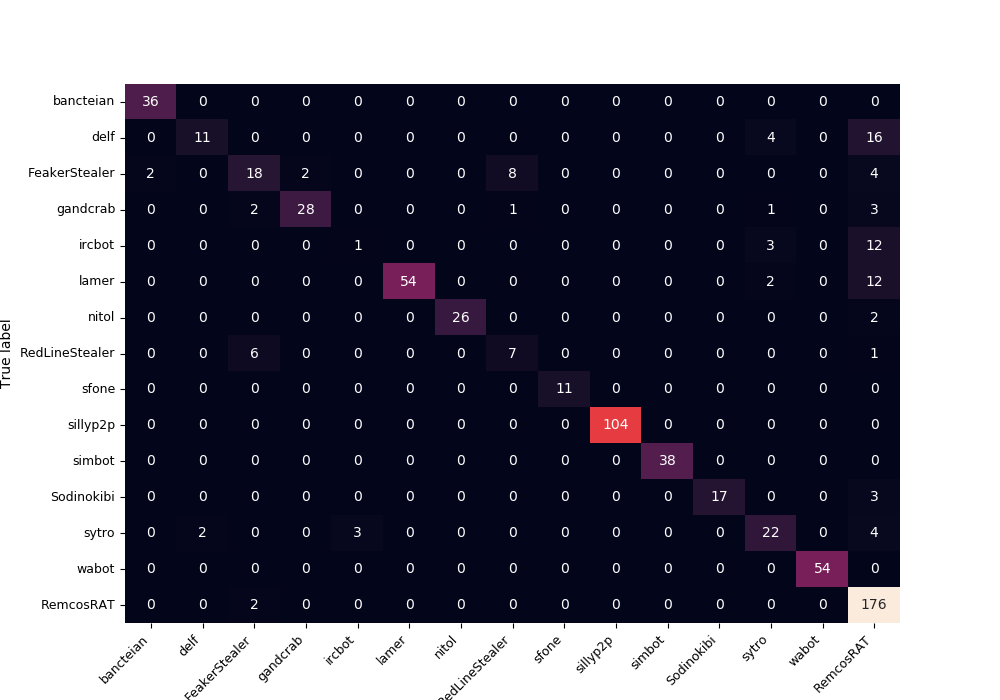}
 \caption{Confusion matrix obtained for SVM classifier and kernel from~\cite{puodzius2021accurate} with \textbf{CBFS-strategy} and \textbf{SCDG-strategy 2-3-4}}
 \label{fig:conf_m_kern}
\end{minipage}\\
\end{figure}

The results are reported in Table~\ref{table:res_gspan}, from which we observe that {\bf CBFS-Strategy} and {\bf BFS-Strategy} generally outperform {\bf CDFS-Strategy}.By inspecting the results, we observed that {\bf BFS-Strategy} ran out of memory for 7 percent of the binaries, thus reducing its performance compared to {\bf CBFS-Strategy}. While \textbf{SCDG-strategy 4} showed improvements with \textbf{SCDG-strategy 5}, it should be noted that \textbf{SCDG-strategy 5} entails significant overhead in SCDG building (up to 100 times slower) and signature size (5 times bigger on average). In general, the best performances were obtained by combining \textbf{SCDG-strategy 2}, \textbf{SCDG-strategy 3}, and \textbf{SCDG-strategy 4}. Upon inspecting the confusion matrix obtained for the best set of parameters in Figure~\ref{fig:conf_m_gspan}, we see a lot of confusion between different classes. This can be explain by plotting similarities between signatures built with Gspan, as illustrated in Figure~\ref{fig:inter_sig_gspan}. One can see that different signatures share important similarities, leading to confusion between different malware families, as illustrated in Figure~\ref{fig:conf_m_gspan}. This problem is directly linked to a problem exposed in Section~\ref{sec:background}, that is, Gspan focus on the computation of the biggest subgraph while neglecting other components.

We now turn to applying kernel from~\cite{puodzius2021accurate}. Table~\ref{table:res_littkernel} shows that overall performance increased compared with Gspan. Moreover, {\bf CBFS-Strategy} and {\bf CDFS-Strategy} outperformed both {\bf BFS-Strategy} and \textbf{SCDG-strategy 2}, \textbf{SCDG-strategy 3}, and \textbf{SCDG-strategy 4} strategies appeared to be more efficient. However, Figure~\ref{fig:conf_m_kern} shows that several families were still indistinguishable. 

\begin{figure}[!ht]
	\centering
	\includegraphics[scale=0.35]{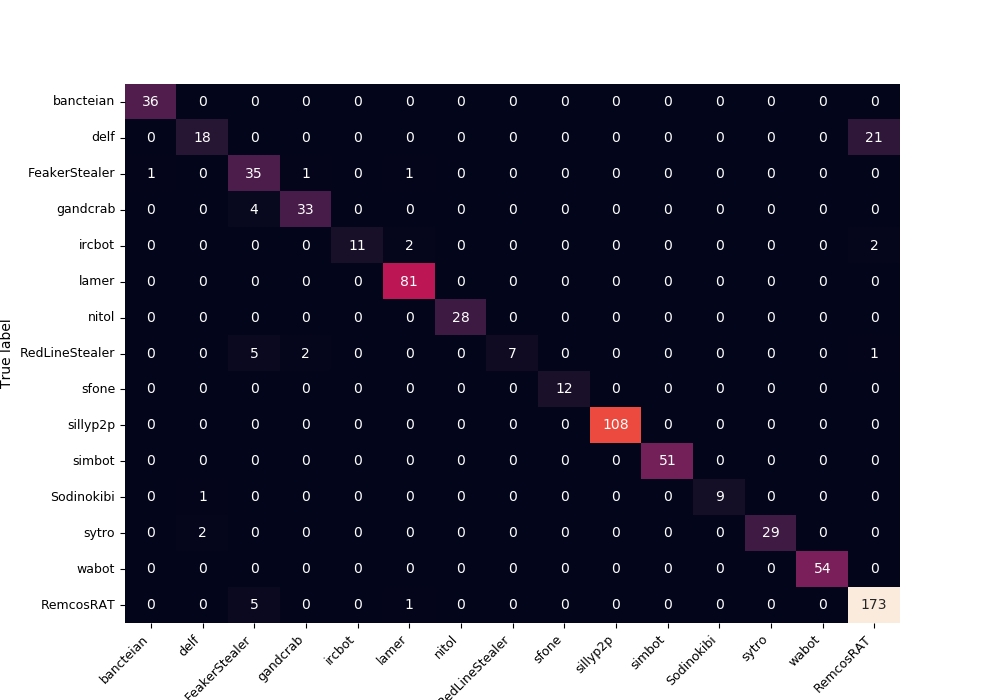}
	\caption{Confusion matrix obtained for WL kernel with \textbf{CDFS-strategy} and \textbf{SCDG-strategy 1-4}. }
	\label{fig:conf_matrix_WL}
\end{figure}

\begin{table}[!htb]
\centering
\begin{tabular}{|lllll|lll|lll|lll|}
\hline
\multicolumn{5}{|l|}{\textbf{SCDG-strategy}} &
  \multicolumn{3}{l|}{{\bf BFS-Strategy}} &
  \multicolumn{3}{l|}{{\bf CBFS-Strategy}} &
  \multicolumn{3}{l|}{{\bf CDFS-Strategy}} \\ \hline
\multicolumn{1}{|l|}{\ 1 \ } &
  \multicolumn{1}{l|}{\ 2 \ } &
  \multicolumn{1}{l|}{\ 3 \ } &
  \multicolumn{1}{l|}{\ 4 \ } &
  \ 5 &
  \multicolumn{1}{l|}{Precision} &
  \multicolumn{1}{l|}{Recall} &
  $F_1$-score &
  \multicolumn{1}{l|}{Precision} &
  \multicolumn{1}{l|}{Recall} &
  $F_1$-score &
  \multicolumn{1}{l|}{Precision} &
  \multicolumn{1}{l|}{Recall} &
  $F_1$-score\\ \hline
\multicolumn{1}{|l|}{\ x} &
  \multicolumn{1}{l|}{} &
  \multicolumn{1}{l|}{\ x} &
  \multicolumn{1}{l|}{\ x} &
   &
  \multicolumn{1}{l|}{0.754} &
  \multicolumn{1}{l|}{0.691} & 0.721
   & 
  \multicolumn{1}{l|}{0.845} &
  \multicolumn{1}{l|}{0.812} & 0.828
   & 
  \multicolumn{1}{l|}{0.833} &
  \multicolumn{1}{l|}{0.787} & 0.8093
   \\ \hline
\multicolumn{1}{|l|}{\ x} &
  \multicolumn{1}{l|}{} &
  \multicolumn{1}{l|}{\ x} &
  \multicolumn{1}{l|}{} &
  \ x &
  \multicolumn{1}{l|}{0.728} &
  \multicolumn{1}{l|}{0.673} & 0.699
   &
  \multicolumn{1}{l|}{0.785} &
  \multicolumn{1}{l|}{0.736} & 0.759
   &
  \multicolumn{1}{l|}{0.77} &
  \multicolumn{1}{l|}{0.714} & 0.741
   \\ \hline
\multicolumn{1}{|l|}{\ x} &
  \multicolumn{1}{l|}{} &
  \multicolumn{1}{l|}{} &
  \multicolumn{1}{l|}{\ x} &
   &
  \multicolumn{1}{l|}{0.742} &
  \multicolumn{1}{l|}{0.71} & 0.725
   &
  \multicolumn{1}{l|}{0.827} &
  \multicolumn{1}{l|}{0.778} & 0.801
   &
  \multicolumn{1}{l|}{0.82} &
  \multicolumn{1}{l|}{0.754} & 0.786
   \\ \hline
\multicolumn{1}{|l|}{\ x} &
  \multicolumn{1}{l|}{} &
  \multicolumn{1}{l|}{} &
  \multicolumn{1}{l|}{} &
  \ x &
  \multicolumn{1}{l|}{0.711} &
  \multicolumn{1}{l|}{0.649} & 0.678
   &
  \multicolumn{1}{l|}{0.771} &
  \multicolumn{1}{l|}{0.703} & 0.735
   &
  \multicolumn{1}{l|}{0.74} &
  \multicolumn{1}{l|}{0.698} & 0.718
   \\ \hline
\multicolumn{1}{|l|}{} &
  \multicolumn{1}{l|}{\ x} &
  \multicolumn{1}{l|}{\ x} &
  \multicolumn{1}{l|}{\ x} &
   &
  \multicolumn{1}{l|}{0.769} &
  \multicolumn{1}{l|}{0.723} & 0.745
   &
  \multicolumn{1}{l|}{0.851} &
  \multicolumn{1}{l|}{0.826} & 0.838
   &
  \multicolumn{1}{l|}{0.847} &
  \multicolumn{1}{l|}{0.813} & 0.829
    \\ \hline
\multicolumn{1}{|l|}{} &
  \multicolumn{1}{l|}{\ x} &
  \multicolumn{1}{l|}{\ x} &
  \multicolumn{1}{l|}{} &
  \ x &
  \multicolumn{1}{l|}{0.738} &
  \multicolumn{1}{l|}{0.645} & 0.688
   &
  \multicolumn{1}{l|}{0.813} &
  \multicolumn{1}{l|}{0.752} & 0.781
   &
  \multicolumn{1}{l|}{0.802} &
  \multicolumn{1}{l|}{0.738} & 0.768
   \\ \hline
\multicolumn{1}{|l|}{} &
  \multicolumn{1}{l|}{\ x} &
  \multicolumn{1}{l|}{} &
  \multicolumn{1}{l|}{\ x} &
   &
  \multicolumn{1}{l|}{0.747} &
  \multicolumn{1}{l|}{0.632} & 0.684
   &
  \multicolumn{1}{l|}{0.798} &
  \multicolumn{1}{l|}{0.747} &
   0.771 & 
  \multicolumn{1}{l|}{0.835} &
  \multicolumn{1}{l|}{0.772} & 0.802
   \\ \hline
\multicolumn{1}{|l|}{} &
  \multicolumn{1}{l|}{\ x} &
  \multicolumn{1}{l|}{} &
  \multicolumn{1}{l|}{} &
  \ x &
  \multicolumn{1}{l|}{0.714} &
  \multicolumn{1}{l|}{0.654} & 0.682
   &
  \multicolumn{1}{l|}{0.781} &
  \multicolumn{1}{l|}{0.733} &
   0.756 &
  \multicolumn{1}{l|}{0.763} &
  \multicolumn{1}{l|}{0.718} & 0.739
   \\ \hline
\end{tabular}
\centering
\caption{Results of SVM classifier and kernel from~\cite{puodzius2021accurate} with different exploration strategies}
\label{table:res_littkernel}
\end{table}
\paragraph{\textbf{Weisfeiler-Lehman kernel}}. Finally, we investigated the SVM classifier with the Weisfeiler-Lehman kernel. The results in Table~\ref{table:res_graphkernel} clearly outperformed the others, reaching an $F_1$-score of 0.929 with \textbf{CDFS-Strategy} and \textbf{SCDG-strategy 1-3-4}. Malware families were better distinguished, as illustrated in the confusion matrix given in Figure~\ref{fig:conf_matrix_WL}. Those observations confirm our supposition exposed in Section~\ref{sec:background}: taking advantage of an SCDG's local structure increases the efficiency of machine learning in malware classification.  Table~\ref{table:classif_result_binaire} shows that those observations also extend to the simpler operation of detecting malware.

\begin{table}[!htb]
\centering
\begin{tabular}{|lllll|lll|lll|lll|}
\hline
\multicolumn{5}{|l|}{\textbf{SCDG-strategy}} &
  \multicolumn{3}{l|}{{\bf BFS-Strategy}} &
  \multicolumn{3}{l|}{{\bf CBFS-Strategy}} &
  \multicolumn{3}{l|}{{\bf CDFS-Strategy}} \\ \hline
\multicolumn{1}{|l|}{\ 1 \ } &
  \multicolumn{1}{l|}{\ 2 \ } &
  \multicolumn{1}{l|}{\ 3 \ } &
  \multicolumn{1}{l|}{\ 4 \ } &
  \ 5 &
  \multicolumn{1}{l|}{Precision} &
  \multicolumn{1}{l|}{Recall} &
  $F_1$-score &
  \multicolumn{1}{l|}{Precision} &
  \multicolumn{1}{l|}{Recall} &
  $F_1$-score &
  \multicolumn{1}{l|}{Precision} &
  \multicolumn{1}{l|}{Recall} &
  $F_1$-score\\ \hline
\multicolumn{1}{|l|}{\ x} &
  \multicolumn{1}{l|}{} &
  \multicolumn{1}{l|}{\ x} &
  \multicolumn{1}{l|}{\ x} &
   &
  \multicolumn{1}{l|}{0.852} &
  \multicolumn{1}{l|}{0.847} & 0.846 
   & 
  \multicolumn{1}{l|}{0.865} &
  \multicolumn{1}{l|}{0.864} &
   0.852 & 
  \multicolumn{1}{l|}{0.936} &
  \multicolumn{1}{l|}{0.931} & 
   \textbf{0.929}\\ \hline
\multicolumn{1}{|l|}{\ x} &
  \multicolumn{1}{l|}{} &
  \multicolumn{1}{l|}{\ x} &
  \multicolumn{1}{l|}{} &
  \ x &
  \multicolumn{1}{l|}{0.832} &
  \multicolumn{1}{l|}{0.824} & 0.827
   &
  \multicolumn{1}{l|}{0.85} &
  \multicolumn{1}{l|}{0.842} & 0.846
   & 
  \multicolumn{1}{l|}{0.915} &
  \multicolumn{1}{l|}{0.91} & 0.912
   \\ \hline
\multicolumn{1}{|l|}{\ x} &
  \multicolumn{1}{l|}{} &
  \multicolumn{1}{l|}{} &
  \multicolumn{1}{l|}{\ x} &
   &
  \multicolumn{1}{l|}{0.895} &
  \multicolumn{1}{l|}{0.891} & 0.892
   &
  \multicolumn{1}{l|}{0.894} &
  \multicolumn{1}{l|}{0.881} &
   0.874 &
  \multicolumn{1}{l|}{0.937} &
  \multicolumn{1}{l|}{0.933} &
  \textbf{0.929 }\\ \hline
\multicolumn{1}{|l|}{\ x} &
  \multicolumn{1}{l|}{} &
  \multicolumn{1}{l|}{} &
  \multicolumn{1}{l|}{} &
  \ x &
  \multicolumn{1}{l|}{0.847} &
  \multicolumn{1}{l|}{0.836} & 0.841
   &
  \multicolumn{1}{l|}{0.86} &
  \multicolumn{1}{l|}{0.851} & 0.855
   &
  \multicolumn{1}{l|}{0.918} &
  \multicolumn{1}{l|}{0.911} & 0.914
   \\ \hline
\multicolumn{1}{|l|}{} &
  \multicolumn{1}{l|}{\ x} &
  \multicolumn{1}{l|}{\ x} &
  \multicolumn{1}{l|}{\ x} &
   &
  \multicolumn{1}{l|}{0.86} &
  \multicolumn{1}{l|}{0.855} & 0.857
   &
    \multicolumn{1}{l|}{0.897} &
  \multicolumn{1}{l|}{0.879} &
   0.867&
  \multicolumn{1}{l|}{0.929} &
  \multicolumn{1}{l|}{0.925} &
   0.924 \\ \hline
\multicolumn{1}{|l|}{} &
  \multicolumn{1}{l|}{\ x} &
  \multicolumn{1}{l|}{\ x} &
  \multicolumn{1}{l|}{} &
  \ x &
  \multicolumn{1}{l|}{0.812} &
  \multicolumn{1}{l|}{0.795} & 0.803
   &
  \multicolumn{1}{l|}{0.867} &
  \multicolumn{1}{l|}{0.862} &0.864
   &
  \multicolumn{1}{l|}{0.885} &
  \multicolumn{1}{l|}{0.877} & 0.881
   \\ \hline
\multicolumn{1}{|l|}{} &
  \multicolumn{1}{l|}{\ x} &
  \multicolumn{1}{l|}{} &
  \multicolumn{1}{l|}{\ x} &
   &
  \multicolumn{1}{l|}{0.895} &
  \multicolumn{1}{l|}{0.891} & 0.891
   &
  \multicolumn{1}{l|}{0.895} &
  \multicolumn{1}{l|}{0.891} &
   0.886 &
  \multicolumn{1}{l|}{0.939} &
  \multicolumn{1}{l|}{0.933} &
   \textbf{0.929}\\ \hline
\multicolumn{1}{|l|}{} &
  \multicolumn{1}{l|}{\ x} &
  \multicolumn{1}{l|}{} &
  \multicolumn{1}{l|}{} &
  \ x &
  \multicolumn{1}{l|}{0.834} &
  \multicolumn{1}{l|}{0.828} & 0.831
   &
  \multicolumn{1}{l|}{0.862} &
  \multicolumn{1}{l|}{0.858} & 0.859
   &
  \multicolumn{1}{l|}{0.891} &
  \multicolumn{1}{l|}{0.887} & 0.888
   \\ \hline
\end{tabular}
\centering
\caption{Results of SVM classifier and the Weisfeiler-Lehman graph kernel with different exploration strategies}
\label{table:res_graphkernel}
\end{table}

\begin{table}[!htb]
\centering
\begin{tabular}{l|l|l|l}
\hline
 Model & Precision & Recall & $F_1$-score \\ \hline
 Gspan & 0.911   & 0.914 & 0.911 \\ 
 SVM kernel from~\cite{puodzius2021accurate} & 0.965 & 0.95  & 0.957 \\ 
 SVM Weisfeiler-Lehman &  0.989 & 0.975 & 0.981\\ \hline
\end{tabular}
\caption{Binary detection results of the different classifiers}
\label{table:classif_result_binaire}
\end{table}

\paragraph{\textbf{Training time}} We also recorded average training time for the three approaches implemented in our toolchain. In general, we observed that  Weisfeiler-Lehman kernel outperforms Gspan by a factor of 15 and Kernel in ~\cite{puodzius2021accurate} by a factor of 10 000. We suspect that the overhead can be explained by the extensive use of pairwise graph mining in the similarity metric presented in Section~\ref{sec:background}. Compare to the Kernel in ~\cite{puodzius2021accurate}, Gspan reduces these number of computation since it first create a signature for each family before comparing those signature with the binary to classify.

These experiments also enable us to draw conclusions with respect to symbolic execution strategies. First, \textbf{SCDG-strategy 1} gives overall better results than \textbf{SCDG-strategy 2} with Weisfeiler-Lehman kernel. That means that the edges added by this strategy provide useful SCDG abstraction information for learning. That is not the case for the other classifier where these information seems to lead to overfitting and \textbf{SCDG-strategy 2} should be preferred.  Moreover, the impact of \textbf{SCDG-strategy 3} varies. While it improves classification for kernels that are based on the biggest common subgraph, its impact when combined with other strategies varies. Finally, observe that while \textbf{SCDG-strategy 5} leads to a considerable overhead, it does not improve performance of any classifier. On the other hand, \textbf{SCDG-strategy 4} leads to more compact signatures, better computation times and good classification performances. Thus, \textbf{SCDG-strategy 4} should be preferred. Regarding exploration strategies, \textbf{BFS-strategy} is generally outperformed by \textbf{CBFS-strategy} while \textbf{CDFS-strategy} outperforms all other exploration strategies when used with the Weisfeiler-Lehman graph kernel.  Generally speaking, the experiments confirm our intuition given in Section~\ref{sec:background} that "better classification is possible with Weisfeiler-Lehman graph kernel."

%% file: Conclusion.tex
\section{Conclusions and Future Work}
\label{sec:conclusion}

We propose an open-source toolchain for malware analysis based on symbolic execution and machine learning~\cite{git_toolchain}. This toolchain exploits SCDGs extracted from malware of the same family to learn the common behavior shared among this family. We also developed and compare several heuristics related to binary exploration and SCDG building in order to improve the use of symbolic execution in the malware analysis domain. Finally, we demonstrate how using the Weisfeiler-Lehman kernel could improve learning from SCDGs compared with other techniques such as Gspan by exploiting information contained in those graphs better. This leads to significant improvements in the malware sample classification and detection. Directions for future work includes new exploration heuristics, such as concolic executions~\cite{Sen15} or smart sampling~\cite{JLS13}. Another objective is to apply our kernel in a non-supervised approach like in ~\cite{puodzius2021accurate}. We are also interested in implementing a distributed version of the toolchain. In this context, the federated learning paradigm should allow us to combine information from different contributors. In addition, we will continue to improve our toolchain with new \textit{simprocedure} and plugin interfaces.